\documentclass[conference]{IEEEtran}
\usepackage[letterpaper, left=0.62in, right=0.62in, bottom=1in, top=0.75in]{geometry}
\IEEEoverridecommandlockouts
\usepackage[binary-units]{siunitx}
\usepackage[caption=false,font=footnotesize]{subfig}
\usepackage{tabularx}
\usepackage{multirow}
\usepackage{color, colortbl}
\usepackage[binary-units]{siunitx}
\DeclareSIUnit{\bps}{bps}
\usepackage{amsmath,amssymb,amsfonts}
\usepackage{cite}
\usepackage{graphicx}
\def\BibTeX{{\rm B\kern-.05em{\sc i\kern-.025em b}\kern-.08em
    T\kern-.1667em\lower.7ex\hbox{E}\kern-.125emX}}
\begin{document}
\title{Orthogonal and Non-Orthogonal Multiple Access for Intelligent Reflection Surface in 6G Systems}

\author{\IEEEauthorblockN{Wei Jiang\IEEEauthorrefmark{1}\IEEEauthorrefmark{2} and Hans D. Schotten\IEEEauthorrefmark{2}\IEEEauthorrefmark{1}}
\IEEEauthorblockA{\IEEEauthorrefmark{1}Intelligent Networking Research Group, German Research Center for Artificial Intelligence (DFKI),  Germany\\
  }
\IEEEauthorblockA{\IEEEauthorrefmark{2}Department of Electrical and Computer Engineering, Technische Universit\"at (TU) Kaiserslautern, Germany\\
 }
}


%


\maketitle

\begin{abstract}
Intelligent reflecting surface (IRS) is envisioned to become a key technology for the upcoming six-generation (6G) wireless system due to its potential of reaping high performance in a power-efficient and cost-efficient way. With its disruptive capability and hardware constraint, the integration of IRS imposes some fundamental particularities on the coordination of multi-user signal transmission. Consequently, the conventional orthogonal and non-orthogonal multiple-access schemes are hard to directly apply because of the joint optimization of active beamforming at the base station and passive reflection at the IRS.  Relying on an alternating optimization method, we develop novel schemes for efficient multiple access in IRS-aided multi-user multi-antenna systems in this paper. Achievable performance in terms of the sum spectral efficiency is theoretically analyzed. A comprehensive comparison of different schemes and configurations is conducted through Monte-Carlo simulations to clarify which scheme is favorable for this emerging 6G paradigm.
\end{abstract}

%
\IEEEpeerreviewmaketitle

\section{Introduction}

Although the fifth-generation (5G) technology is still on its way to being deployed worldwide, both academia and industry have already shifted their focus on the sixth-generation (6G) technology and enthusiastically initiated many pioneering research programs \cite{Ref_jiang2021kickoff}. To support disruptive use cases beyond 2030 such as Metaverse, holographic telepresence, and digital twin, 6G needs to meet more stringent performance requirements than its predecessor, e.g., a peak rate of terabits-per-second, ultra-massive connectivity, and extreme reliability \cite{Ref_jiang2021road}. Traditionally, three major approaches, i.e.,  \textit{(1) Deploying Dense and Heterogeneous Networks},  \textit{(2) Installing Massive Antennas for Extreme Spectral Efficiency}, and \textit{(3) Enlarging Bandwidth} can effectively improve coverage and capacity. Nevertheless, these approaches incur high capital and operational expenditures, unaffordable energy consumption, and severe network interference. Given these limitations, further evolving along the old track is hard to fully achieve stringent 6G requirements. Therefore, it is highly desirable to develop a revolutionary technology to realize sustainable capacity and performance growth with affordable cost, low complexity, and efficient energy consumption.

Recently, a disruptive technique referred to as intelligent reflecting surface (IRS) has drawn much attention from both academia and industry due to its potential to simultaneously meet the aforementioned demands \cite{Ref_renzo2020smart}. Particularly, IRS is a planar meta-surface composed of a large number of reflection elements, each of which can independently induce a phase shift and amplitude attenuation (collectively termed as reflection coefficient) to an impinging electromagnetic wave \cite{Ref_wu2019intelligent}. These elements thereby collaboratively achieve a smart propagation environment for signal amplification or interference suppression \cite{Ref_renzo2020reconfigurable}. Because the reflection elements are passive, cheap, and lightweight, the IRS is a green and cost-efficient technology. Therefore, it is recognized as a possible enabler for the forthcoming 6G system \cite{Ref_jiang2021road}. It also exhibits great potential to be transparently installed in legacy networks for performance enhancement. A large and growing body of literature have investigated different aspects for building an IRS-aided wireless system, i.e., reflection optimization design \cite{Ref_wu2019intelligent}, cascaded channel estimation \cite{Ref_wang2020channel}, practical constraints \cite{Ref_zhi2021uplink}, discrete phase shifts \cite{Ref_wu2020beamforming}, and the interplay of IRS with other wireless technologies, e.g.,  orthogonal frequency-division multiplexing (OFDM) \cite{Ref_zheng2020intelligent}, multi-input multi-output (MIMO) \cite{Ref_hu2018beyond}, hybrid beamforming \cite{Ref_di2020hybrid}, and Terahertz communications \cite{Ref_jiang2022dualbeam}.

The majority of prior works merely consider point-to-point IRS communications, including a single user, for ease of analysis. Nevertheless, a practical wireless system needs to accommodate many users simultaneously, raising the problem of multiple access.  Due to its disruptive capability in smartly reconfiguring the wireless propagation environment, as well as hardware constraints,  the integration of IRS brings some fundamental particularities on the coordination of multi-user signal transmission. For instance, the lack of frequency-selective reflection, namely the phase shift of each reflection element cannot be different across frequency subchannels, leads to the performance loss of frequency-division approaches. In addition, active beamforming at the base station and passive reflection at the IRS are coupled, and the joint optimization is required.
Consequently, the conventional orthogonal and non-orthogonal multiple-access schemes are hard to directly apply.

Relying on alternating optimization, we develop novel multiple-access methods by upgrading the conventional ones, including time-division multiple access (TDMA), frequency-division multiple access (FDMA), and non-orthogonal multiple access (NOMA), to adapt to IRS-aided multi-user multi-antenna systems. To differentiate with the originals, we hereinafter name these schemes TDMA-IRS, FDMA-IRS, and NOMA-IRS, respectively. Achievable performance in terms of the sum spectral efficiency is theoretically analyzed. A comprehensive comparison of different schemes and configurations is conducted through Monte-Carlo simulations to clarify which scheme is favorable for this emerging 6G paradigm.
The rest of this paper is organized as follows: Section II introduces the system model. Section III develops and analyzes orthogonal and non-orthogonal multiple access for IRS. Simulation setup and numerical results are demonstrated in Section IV. Finally, the conclusions are drawn in Section V.

\section{System Model}
\begin{figure}[!t]
    \centering
    \includegraphics[width=0.42\textwidth]{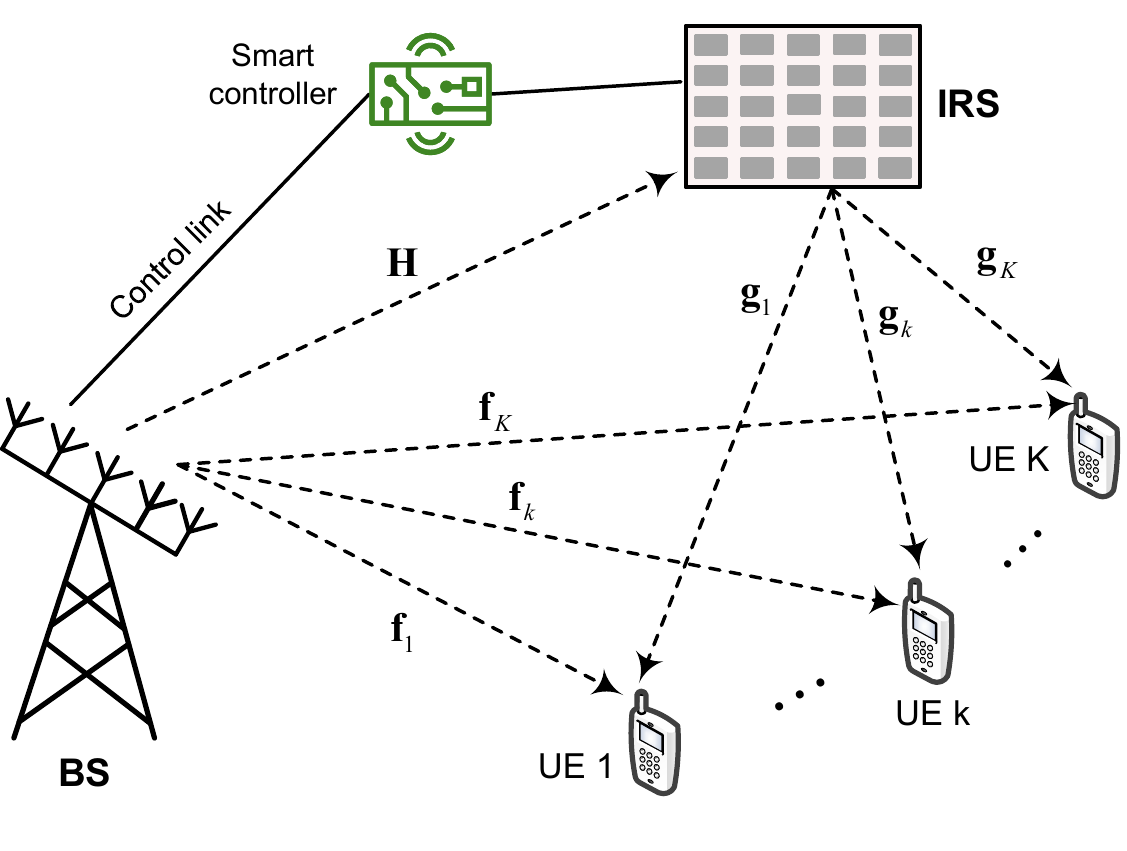}
    \caption{Schematic diagram of an IRS-aided multi-user MIMO system, consisting of a multi-antenna BS, $K$ single-antenna UE, and a reflecting surface with $N$ IRS elements.  }
    \label{diagram:system}
\end{figure}

As illustrated in \figurename \ref{diagram:system}, this paper considers the downlink of an IRS-assisted multi-user MIMO communications system, where an intelligent surface with $N$ reflecting elements is deployed to assist the transmission from an $N_b$-antenna base station (BS) to $K$ single-antenna user equipment (UE). The IRS is a passive device, where time-division duplexing (TDD) is usually adopted to simplify channel estimation. The users send pilot signals in the uplink training so that the BS can estimate the instantaneous channel state information (CSI), which is used for optimizing the downlink data transmission due to channel reciprocity. For ease of illustration, the analysis hereinafter is conducted under the assumption that the BS perfectly knows the CSI of all involved channels, as most prior works \cite{Ref_wu2019intelligent, Ref_renzo2020reconfigurable, Ref_renzo2020smart}. Without losing generality, it assumes that the system experiences flat-fading channels since a frequency-selective channel can be treated as a set of flat-fading channels through OFDM \cite{Ref_jiang2016ofdm}.
Consequently, we write
\begin{align}
    \mathbf{f}_{k}=\Bigl[f_{k1},f_{k2},\ldots,f_{kN_b}\Bigr]^T
\end{align}
to denote the $N_b\times 1$ channel vector between the BS and the $k^{th}$ UE, and
\begin{align}
    \mathbf{g}_{k}=\Bigl[g_{k1},g_{k2},\ldots,g_{kN}\Bigr]^T
\end{align}
to denote the $N\times 1$ channel vector between the IRS and UE $k$. Denoting  the channel vector from the BS to the $n^{th}$ reflecting element by $\mathbf{h}_{n}=[h_{n1},h_{n2},\ldots,h_{nN_b}]^T$, the channel matrix from the BS to the IRS is expressed as $\mathbf{H}\in \mathbb{C}^{N\times N_b}$, where the $n^{th}$ row of $\mathbf{H}$ equals to $\mathbf{h}_n^T$.

Since the line-of-sight (LOS) paths from either the BS or the IRS to UEs may be blocked, the corresponding small-scale fading follows Rayleigh distribution. In other words, $f$ and $g$ are circularly symmetric complex Gaussian random variables, denoted by $f\sim \mathcal{CN}(0,\sigma_f^2)$ and $g\sim \mathcal{CN}(0,\sigma_g^2)$, respectively.
The variances $\sigma_f^2$ and $\sigma_g^2$ mean distance-dependent large-scale fading. It can be computed by
$10^\frac{\mathcal{P}+\mathcal{S}}{10}$, where $\mathcal{P}$ is distance-dependent path loss, $\mathcal{S}$ stands for \textit{log-normal} shadowing denoted by $\mathcal{S}\sim \mathcal{N}(0,\sigma_{sd}^2)$. As  \cite{Ref_jiang2021cellfree},  this paper adopts the COST-Hata model to calculate the path loss for the UEs, i.e.,
\begin{equation} \label{eqn:CostHataModel}
    \mathcal{P}=
\begin{cases}
-\mathcal{P}_0-35\log_{10}(x), &  x>x_1 \\
-\mathcal{P}_0-15\log_{10}(x_1)-20\log_{10}(x), &  x_0<x\leqslant x_1 \\
-\mathcal{P}_0-15\log_{10}(x_1)-20\log_{10}(x_0), &  x\leqslant x_0
\end{cases},
\end{equation}
where $x$ is the link distance,  $x_0$ and $x_1$ represent the break points of the three-slope model, and $\mathcal{P}_0$ is a constant.
In contrast, a favourable location  is deliberately selected for the IRS to exploit an LOS path to the fixed BS without any blockage, resulting in Rician fading, i.e.,
\begin{equation}\label{EQNIRQ_LSFadingdirect}
    \mathbf{H}=\sqrt{\frac{\Gamma\sigma_h^2}{\Gamma+1}}\mathbf{H}_{LOS} + \sqrt{\frac{\sigma_h^2}{\Gamma+1}}\mathbf{H}_{NLOS}
\end{equation}
with the Rician factor $\Gamma$, the LOS component $\mathbf{H}_{LOS}$,  the multipath component $\mathbf{H}_{NLOS}$ consisting of independent entries that follow $\mathcal{CN}(0,1)$, and the BS-IRS path loss $\sigma_h^2$. Consider a macro-cell scenario with far-field assumption, where there is no reflector surrounding the BS and IRS, the signal transmission more likes free-space propagation, which can be computed by
\begin{equation}
    \sigma_h^2=\frac{L_0}{x^{-\alpha}},
\end{equation}
where $L_0$ stands for the path loss at the reference distance of \SI{1}{\meter}, and $\alpha$ means the path loss exponent.

A smart controller of the IRS is connected to the BS with a wired or wireless link. It is responsible for adaptively adjusting the phase shift of each reflecting element in terms of the acquired CSI through periodic channel estimation \cite{Ref_wang2020channel}. The signal reflection of a typical element $n$ for user $k$ is mathematically modeled by a reflection coefficient $\epsilon_{kn}=a_{kn} e^{j\phi_{kn}}$, where $\phi_{kn}\in [0,2\pi)$ denotes an induced phase shift, and $a_{kn}\in [0,1]$ stands for amplitude attenuation. As mentioned by \cite{Ref_wu2019intelligent}, $a_{kn}=1$, $\forall n, k$ is the optimal attenuation that maximizes the strength of the received signal and simplifies the implementation complexity. Hence, the reflection optimization only focuses on the phase shifts $\phi_{kn}$, $\forall n,k$. By ignoring hardware impairments such as quantified phase shifts \cite{Ref_wu2020beamforming} and phase noise \cite{Ref_jiang2022impact}, the $k^{th}$ UE observes the received signal
\begin{equation} \label{eqn_systemModel}
    r_k=\sqrt{P_d}\Biggl( \sum_{n=1}^N g_{kn} e^{j\phi_{kn}} \mathbf{h}_{n}^T + \mathbf{f}_{k}^T \Biggr) \mathbf{s} + n_k,
\end{equation}
where $\mathbf{s}$ stands for the vector of transmitted symbols, $P_d$ represents the BS power constraint, $n_k$ denotes additive white Gaussian noise (AWGN) with zero mean and variance $\sigma_n^2$, i.e., $n_k\sim \mathcal{CN}(0,\sigma_n^2)$. Define a diagonal phase-shift matrix as $\boldsymbol{\Theta}_k=\mathrm{diag}\{e^{j\phi_{1k}},\ldots,e^{j\phi_{Nk}}\}$, \eqref{eqn_systemModel} can be rewritten in matrix form as
\begin{equation} \label{EQN_IRS_RxSignal_Matrix}
    r_k= \sqrt{P_d}\Bigl(\mathbf{g}_k^T \boldsymbol{\Theta}_k \mathbf{H} +\mathbf{f}_k^T\Bigr)\mathbf{s} +n_k.
\end{equation}

\section{Multiple Access for IRS}
In this section, we elaborate the fundamentals of TDMA-IRS, FDMA-IRS, and NOMA-IRS, respectively, where an alternating method is introduced to jointly optimize active beamforming at the BS and passive reflection at the IRS. Their closed-form expressions of the sum spectral efficiency are derived.
\subsection{TDMA-IRS}  This scheme divides the signaling dimensions along the time axis into orthogonal portions  called time slots. Each user transmits over the entire bandwidth but cyclically accesses its assigned slot. It implies non-continuous transmission, which simplifies the system design since some processing such as channel estimation can be performed during the time slots of other users. Another advantage is that TDMA is able to assign multiple time slots for a single user, increasing system flexibility. Mathematically, a radio frame is orthogonally divided into $K$ time slots, where the CSI keeps constant. Using the round-robin scheduling, the BS applies linear beamforming $\mathbf{w}_k\in \mathbb{C}^{N_b\times 1}$, where $\|\mathbf{w}_k\|^2\leqslant 1$, to transmit the signal intended for a general user $k$ at the $k^{th}$ slot. The information-bearing symbol $s_k$ is zero mean and unit-variance, i.e., $\mathbb{E}\left[|s_k|^2\right]=1$.  Substituting $\mathbf{s}=\mathbf{w}_k s_k$ into \eqref{EQN_IRS_RxSignal_Matrix}, we obtain
\begin{equation}
    r_k= \sqrt{P_d}\Bigl(\mathbf{g}_k^T \boldsymbol{\Theta}_k \mathbf{H} +\mathbf{f}_k^T\Bigr)\mathbf{w}_k s_k +n_k.
\end{equation}

By jointly optimizing active beamforming $\mathbf{w}_k$ and reflection $\boldsymbol{\Theta}_k$, the instantaneous SNR of user $k$, i.e.,
\begin{equation} \label{IRS_EQN_spectralEfficiency}
    \gamma_k=\frac{P_d \Bigl|\bigl(\mathbf{g}_k^T \boldsymbol{\Theta}_k \mathbf{H} +\mathbf{f}_k^T\bigr)\mathbf{w}_k\Bigr|^2 }{\sigma_n^2}
\end{equation}
can be maximized, formulating the following optimization
\begin{equation}
\begin{aligned} \label{eqnIRS:optimizationMRTvector}
\max_{\boldsymbol{\Theta}_k,\:\mathbf{w}_k}\quad &  \biggl|\Bigl(\mathbf{g}_k^T \boldsymbol{\Theta}_k \mathbf{H} +\mathbf{f}_k^T\Bigr)\mathbf{w}_k\biggr|^2\\
\textrm{s.t.} \quad & \|\mathbf{w}_k\|^2\leqslant 1\\
  \quad & \phi_{nk}\in [0,2\pi), \: \forall n=1,\ldots,N, \forall k=1,\ldots,K,
\end{aligned}
\end{equation}
which is non-convex because the objective function is not jointly concave with respect to $\boldsymbol{\Theta}_k$ and $\mathbf{w}_k$. To solve this problem, we can apply alternating optimization that alternately optimizes $\boldsymbol{\Theta}_k$ and $\mathbf{w}_k$ in an iterative manner \cite{Ref_wu2019intelligent}.
Without loss of generality, the maximal-ratio transmission (MRT) for the direct link can be applied as the initial value of the transmit vector, i.e., $\mathbf{w}_{k}^{(0)}=\mathbf{f}_{k}^*/\|\mathbf{f}_{k}\|$.
Thus, \eqref{eqnIRS:optimizationMRTvector} is simplified to
\begin{equation}  \label{eqnIRS:optimAO}
\begin{aligned} \max_{\boldsymbol{\Theta}_k}\quad &  \biggl|\Bigl(\mathbf{g}_k^T \boldsymbol{\Theta}_k \mathbf{H} +\mathbf{f}_{k}^T\Bigr)\mathbf{w}_k^{(0)}\biggr|^2\\
\textrm{s.t.}  \quad & \phi_{nk}\in [0,2\pi), \: \forall n=1,\ldots,N, \forall k=1,\ldots,K.
\end{aligned}
\end{equation}
The objective function is still non-convex but it enables a closed-form solution through applying the well-known triangle inequality
\begin{equation}
    \biggl|\Bigl(\mathbf{g}_k^T \boldsymbol{\Theta}_k \mathbf{H} +\mathbf{f}_{k}^T\Bigr)\mathbf{w}_k^{(0)}\biggr| \leqslant \biggl|\mathbf{g}_k^T \boldsymbol{\Theta}_k \mathbf{H} \mathbf{w}_k^{(0)}\biggr| +\biggl|\mathbf{f}_{k}^T\mathbf{w}_k^{(0)}\biggr|.
\end{equation}
The equality achieves if and only if
\begin{equation}
    \arg\left (\mathbf{g}_k^T \boldsymbol{\Theta}_k \mathbf{H} \mathbf{w}_k^{(0)}\right)= \arg\left(\mathbf{f}_{k}^T\mathbf{w}_k^{(0)}\right)\triangleq \varphi_{0k},
\end{equation}
where $\arg(\cdot)$ denotes the angle of a complex vector or scalar.

Define $\mathbf{q}_k=\left[q_{1k},q_{2k},\ldots,q_{Nk}\right]^H$ with $q_{nk}=e^{j\phi_{nk}}$ and  $\boldsymbol{\chi}_k=\mathrm{diag}(\mathbf{g}_k^T)\mathbf{H}\mathbf{w}_k^{(0)}\in \mathbb{C}^{N\times 1}$, we have $\mathbf{g}_k^T \boldsymbol{\Theta}_k \mathbf{H} \mathbf{w}_k^{(0)}=\mathbf{q}_k^H\boldsymbol{\chi}_k\in \mathbb{C} $.
Ignore the constant term $\bigl|\mathbf{f}_{k}^T\mathbf{w}_k^{(0)}\bigr|$, \eqref{eqnIRS:optimAO} is transformed to
\begin{equation}  \label{eqnIRS:optimizationQ}
\begin{aligned} \max_{\boldsymbol{\mathbf{q}_k}}\quad &  \Bigl|\mathbf{q}_k^H\boldsymbol{\chi}_k\Bigl|\\
\textrm{s.t.}  \quad & |q_{nk}|=1, \: \forall n=1,\ldots,N, \forall k=1,\ldots,K,\\
  \quad & \arg(\mathbf{q}_k^H\boldsymbol{\chi}_k)=\varphi_{0k}.
\end{aligned}
\end{equation}
The solution for \eqref{eqnIRS:optimizationQ} can be derived as
\begin{equation} \label{eqnIRScomplexityQ}
    \mathbf{q}^{(1)}_k=e^{j\left(\varphi_{0k}-\arg(\boldsymbol{\chi}_k)\right)}=e^{j\left(\varphi_{0k}-\arg\left( \mathrm{diag}(\mathbf{g}_k^T)\mathbf{H}\mathbf{w}_k^{(0)}\right)\right)}.
\end{equation}
Accordingly,
\begin{align} \nonumber \label{IRSeqnOptimalShift}
    \phi_{nk}^{(1)}&=\varphi_{0k}-\arg\left(g_{nk}\mathbf{h}_{n}^T\mathbf{w}_k^{(0)}\right)\\&=\varphi_{0k}-\arg\left(g_{nk}\right)-\arg\left(\mathbf{h}_{n}^T\mathbf{w}_k^{(0)}\right),
\end{align}
where $\mathbf{h}_{n}^T\mathbf{w}_k^{(0)}\in \mathbb{C} $ can be regarded as an effective SISO channel perceived by the $n^{th}$ reflecting element, combining the effects of transmit beamforming $ \mathbf{w}_k^{(0)}$ and channel response $\mathbf{h}_{n}$. In this regard, \eqref{IRSeqnOptimalShift} implies that an IRS reflector should be tuned such that the phase of the reflected signal through the cascaded link is compensated, and the residual phase is aligned with that of the signal over the direct link, to achieve coherent combining at the receiver.
Once the reflecting phases at the first iteration, i.e., $    \boldsymbol{\Theta}^{(1)}_k=\mathrm{diag}\left\{e^{j\phi_{1k}^{(1)}},e^{j\phi_{2k}^{(1)}},\ldots,e^{j\phi_{Nk}^{(1)}}\right\}$ are determined, the optimization is alternated to update $\mathbf{w}_k$. The BS can apply MRT to maximize the strength of a desired signal, resulting in
\begin{equation}  \label{EQN_IRS_TXBF}
    \mathbf{w}_k^{(1)} = \frac{\left(\mathbf{g}_k^T \boldsymbol{\Theta}^{(1)}_k \mathbf{H} +\mathbf{f}_{k}^T\right)^H}{\left\|\mathbf{g}_k^T \boldsymbol{\Theta}_k^{(1)} \mathbf{H} +\mathbf{f}_{k}^T\right\|}.
\end{equation}

After the completion of the first iteration, the BS gets $\boldsymbol{\Theta}^{(1)}_k$ and $\mathbf{w}_k^{(1)}$, which serve as the initial input for the second iteration to derive $\boldsymbol{\Theta}^{(2)}_k$ and $\mathbf{w}_k^{(2)}$.
This process iterates until the convergence is achieved with the optimal beamformer $\mathbf{w}_k^{\star}$ and  optimal reflection $\boldsymbol{\Theta}_k^{\star}$. Substituting $\mathbf{w}_k^{\star}$ and  $\boldsymbol{\Theta}_k^{\star}$ into \eqref{IRS_EQN_spectralEfficiency}, we can derive the achievable spectral efficiency of user $k$.
Thereby, the sum rate of the TDMA-IRS system can be computed by
\begin{equation} \label{IRS_EQN_TDMA_SE}
    C_{tdma}=\sum_{k=1}^K\frac{1}{K}\log\left(1+\frac{P_d \Bigl|\bigl(\mathbf{g}_k^T \boldsymbol{\Theta}_k^\star \mathbf{H} +\mathbf{f}_k^T\bigr)\mathbf{w}_k^\star \Bigr|^2 }{\sigma_n^2} \right).
\end{equation}

\subsection{FDMA-IRS} In FDMA, the system bandwidth is divided along the frequency axis into $K$ orthogonal subchannels. Each user occupies a dedicated subchannel over the entire time. The BS employs linear beamforming $\mathbf{w}_k$ to transmit $s_k$ over the $k^{th}$ subchannel with equally-allocated transmit power $P_d/K$. Thus, the achievable spectral efficiency of user $k$ is
\begin{equation}
    R_k=\frac{1}{K}\log\left(1+\frac{P_d/K \left|\bigl(\mathbf{g}_k^T \boldsymbol{\Theta}_k \mathbf{H} +\mathbf{f}_k^T\bigr)\mathbf{w}_k\right|^2 }{\sigma_n^2/K} \right).
\end{equation}
In contrast to TDMA, where the IRS phase shifts can be dynamically adjusted in different slots,  the surface can be optimized only for a particular user, whereas other users suffer from phase-unaligned reflection. That is because the hardware limitation of IRS passive elements, which can be fabricated in \textit{time-selective} rather than \textit{frequency-selective}.

Without losing generality, we suppose the FDMA-IRS system optimizes the IRS to aid  the signal transmission of user $\hat{k}$, the optimal parameters $\boldsymbol{\Theta}^{\star}_{\hat{k}}$ and $\mathbf{w}_{\hat{k}}^{\star}$ can be derived using the same alternating optimization as that of TDMA-IRS. Once the phase shifts of the surface are completely adjusted for $\hat{k}$,  what the remaining $K-1$ users, denoted by $\{i|i=1,2,\ldots,K, \:i \neq \hat{k}\}$, can do is to realize partial optimization (instead of joint optimization) by updating their respective active beamforming based on the combined channel gain $\mathbf{g}_{i}^T \boldsymbol{\Theta}^{\star}_{\hat{k}} \mathbf{H} +\mathbf{f}_{i}^T$. For user $i$, the beamformer can be optimized as
\begin{equation}  \label{EQN_IRS_matchedFilter}
    \mathbf{w}_{i}^{\star} = \frac{\Bigl(\mathbf{g}_{i}^T \boldsymbol{\Theta}^{\star}_{\hat{k}} \mathbf{H} +\mathbf{f}_{i}^T\Bigr)^H}{\Bigl\|\mathbf{g}_{i}^T \boldsymbol{\Theta}_{\hat{k}}^{\star} \mathbf{H} +\mathbf{f}_{i}^T\Bigr\|}.
\end{equation}
 Then, the sum rate of the FDMA-IRS system is calculated by
 \begin{align}
     C_{fdma}&= \frac{1}{K}\log\left(1+\frac{P_d \Bigl|\bigl(\mathbf{g}_{\hat{k}}^T \boldsymbol{\Theta}_{\hat{k}}^\star \mathbf{H} +\mathbf{f}_{\hat{k}}^T\bigr)\mathbf{w}_{\hat{k}}^\star\Bigr|^2 }{\sigma_n^2} \right)  \\ \nonumber
     &+\sum_{i} \frac{1}{K}\log\left(1+\frac{P_d \Bigl|\bigl(\mathbf{g}_{i}^T \boldsymbol{\Theta}_{\hat{k}}^\star \mathbf{H} +\mathbf{f}_{i}^T\bigr)\mathbf{w}_{i}^\star\Bigr|^2 }{\sigma_n^2} \right).
  \end{align}

\subsection{NOMA-IRS}
Although the inter-user interference among orthogonally multiplexed users is mitigated to facilitate low-complexity multi-user detection at the receiver, it is widely recognized that orthogonal multiple access (OMA) cannot achieve the sum-rate capacity of a multi-user wireless system. Superposition coding and successive interference cancellation (SIC) make it possible to reuse each orthogonal resource unit by more than one user. At the transmitter, all information symbols are superimposed into a single waveform, while the SIC at the receiver decodes the signals iteratively until it gets the desired signal.

\begin{figure}[!t]
    \centering
    \includegraphics[width=0.42\textwidth]{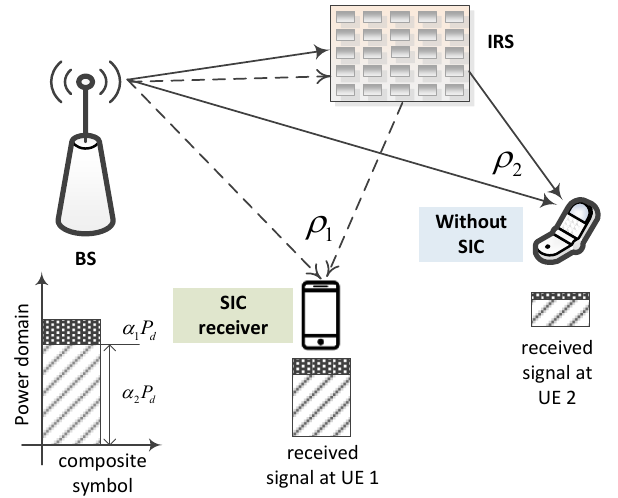}
    \caption{Illustration of a NOMA-IRS system consisting of a BS, a surface, a far user, and a near user.  }
    \label{diagram:NOMA}
\end{figure}

Mathematically, the BS superimposes $K$ information-bearing symbols into a composite waveform
\begin{equation} \label{EQNIRQ_compositeSignal}
    \mathbf{s}=\sum_{k=1}^K \sqrt{\alpha_k } \mathbf{w}_{k} s_k,
\end{equation}
where  $\alpha_k$ represents the power allocation coefficient subjecting to $\sum_{k=1}^K\alpha_k\leqslant 1$. The challenge is to decide how to allocate the power among the users, which is critical for interference cancellation at the receiver. That is why NOMA is regarded as a kind of power-domain multiple access. Generally, more power is  allocated to the users with smaller channel gain, e.g., located farther from the BS, to improve the received SNR, so that high detection reliability can be guaranteed. Despite of less power assigned to a user with a stronger channel gain, e.g., close to the BS, it is capable of detecting its signal correctly with reasonable SNR. As an example, the illustration of a NOMA-IRS system consisting of a BS, a surface, and $K=2$ users are given in \figurename \ref{diagram:NOMA}.

Substitute \eqref{EQNIRQ_compositeSignal} into \eqref{EQN_IRS_RxSignal_Matrix} to yield the observation of user $k$ as
\begin{align} \nonumber
    r_k&= \sqrt{P_d}\Bigl(\mathbf{g}_k^T \boldsymbol{\Theta}_k \mathbf{H} +\mathbf{f}_k^T\Bigr) \sum_{k'=1}^K \sqrt{\alpha_{k'}} \mathbf{w}_{k'}s_{k'} +n_k\\
    &= \underbrace{\sqrt{\alpha_{k}P_d}\Bigl(\mathbf{g}_k^T \boldsymbol{\Theta}_k \mathbf{H} +\mathbf{f}_k^T\Bigr) \mathbf{w}_{k}s_{k}}_{\text{Desired\:signal}}\\ \nonumber
    &+ \underbrace{\sqrt{P_d}\Bigl(\mathbf{g}_k^T \boldsymbol{\Theta}_k \mathbf{H} +\mathbf{f}_k^T\Bigr) \sum_{k'=1,k'\neq k}^K \sqrt{\alpha_{k'}} \mathbf{w}_{k'}s_{k'}}_{\text{Multi-user\:interference}}+n_k.
\end{align}

Due to the hardware limit, the IRS can only assist one user while other users have to share the common phase shifts that are not favorable for them. As FDMA-IRS, we suppose the system optimizes the IRS to aid the signal transmission of user $\hat{k}$.
The optimal parameters $\boldsymbol{\Theta}^{\star}_{\hat{k}}$ and $\mathbf{w}_{\hat{k}}^{\star}$ can be derived using the alternating optimization. Once the phase shifts of the surface are completely adjusted for $\hat{k}$,  a user $k \neq \hat{k}$ can partially optimize its transmission by deriving its active beamforming given the combined channel gain $\mathbf{g}_{k}^T \boldsymbol{\Theta}^{\star}_{\hat{k}} \mathbf{H} +\mathbf{f}_{k}^T$. Similar to \eqref{EQN_IRS_matchedFilter}, the beamformer for user $k$ is figured out as \begin{equation}
    \mathbf{w}_{k}^{\star} = \frac{\Bigl(\mathbf{g}_{k}^T \boldsymbol{\Theta}^{\star}_{\hat{k}} \mathbf{H} +\mathbf{f}_{k}^T\Bigr)^H}{\Bigl\|\mathbf{g}_{k}^T \boldsymbol{\Theta}_{\hat{k}}^{\star} \mathbf{H} +\mathbf{f}_{k}^T\Bigr\|}.
\end{equation}

The same signal $\mathbf{x}$ that contains all information symbols is delivered to all users. The optimal order of interference cancellation is detecting the user with the most power allocation (the weakest channel gain) to the user with the least power allocation (the strongest channel gain). We write $\rho_k=(\mathbf{g}_k^T \boldsymbol{\Theta}^{\star}_{\hat{k}} \mathbf{H} +\mathbf{f}_k^T)\mathbf{w}_k^\star$, $\forall k$ to denote the effective gain of the combined channel for user $k$.
Without loss of generality, assume that user $1$ has the largest combined channel gain, and user $K$ is the weakest, i.e.,
\begin{equation}
   \| \rho_1 \|^2 \geqslant \|\rho_2\|^2 \geqslant \ldots \geqslant \|\rho_K\|^2.
\end{equation}
 With this order, each NOMA-IRS user decodes $s_K$ first, and then subtracts its resultant component from the received signal. As a result, a typical user $k$ after the first SIC iteration gets
\begin{equation}
    \tilde{r}_k = r_k - \rho_k \sqrt{\alpha_K P_d}  s_K = \rho_k\sum_{k=1}^{K-1} \sqrt{\alpha_k P_d} s_k+n_k,
\end{equation}
assuming error-free detection and perfect channel knowledge.
In the second iteration, the user decodes $s_{K-1}$ using the remaining signal $ \tilde{r}_k$. The cancellation iterates until each user gets the symbol intended for it. Particularly, the weakest user decodes its own signal directly without successive interference cancellation since it is allocated the most power. Treating the multi-user interference as noise, the SNR for user $K$ can be written as
\begin{equation}
    \gamma_K=\frac{\|\rho_K \|^2 \alpha_K P_d}{\|\rho_K \|^2\sum_{k=1}^{K-1}\alpha_k P_d + \sigma_n^2 }.
\end{equation}
In general, user $k$ successfully cancels the signals from user $k+1$ to $K$ but suffering from the interference from user $1$ to $k-1$. Consequently, the received SNR for user $k$ is
\begin{equation}
    \gamma_k=\frac{\|\rho_k\|^2 \alpha_k P_d}{ \|\rho_k\|^2 \sum_{k'=1}^{k-1}\alpha_{k'} P_d + \sigma_n^2 },
\end{equation}
resulting in the achievable rate of $
    R_k= \log \left ( 1+\gamma_k \right)$.
The sum rate of NOMA-IRS is computed by
\begin{equation}
    C_{noma}=\sum_{k=1}^K \log \left ( 1+\frac{\|\rho_k\|^2\alpha_k P_d}{\|\rho_k\|^2 \sum_{k'=1}^{k-1}\alpha_{k'} P_d + \sigma_n^2 }   \right).
\end{equation}


\section{Simulation Results}

\begin{figure}[!t]
    \centering
    \includegraphics[width=0.42\textwidth]{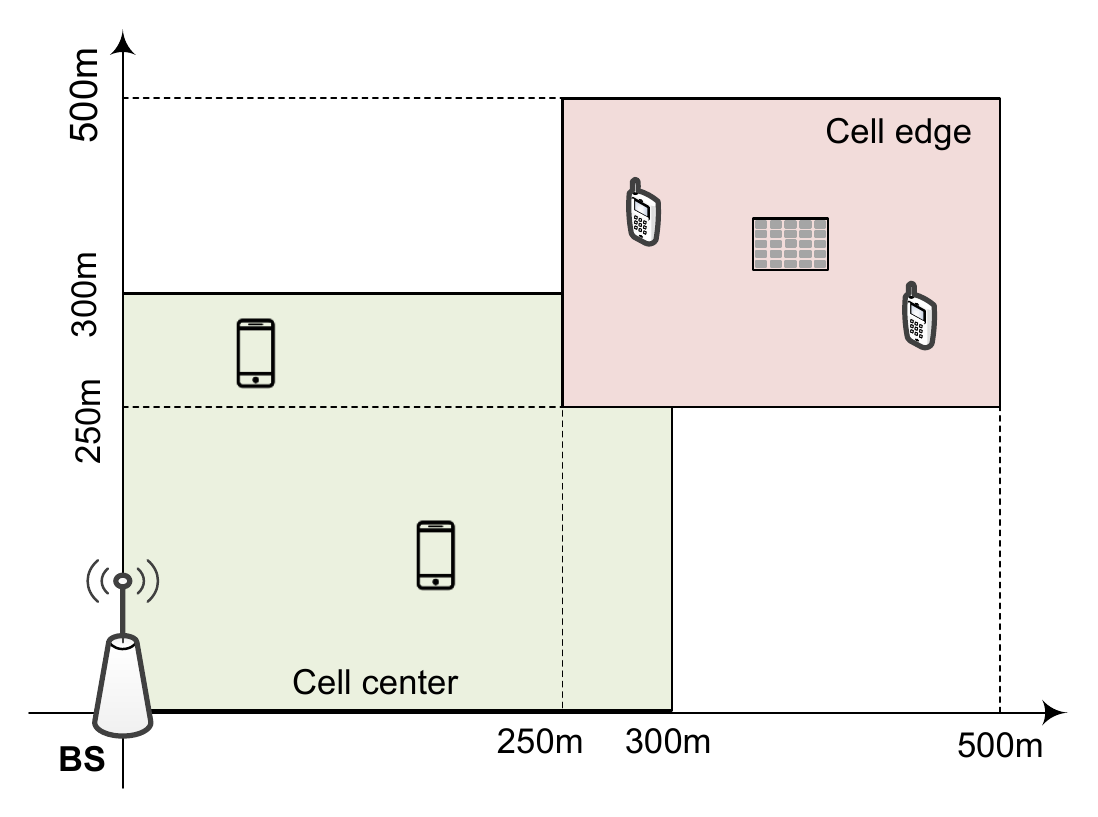}
    \caption{Simulation scenario of a multi-user IRS system, where the cell coverage is comprised of a cell-center area and a cell-edge area. }
    \label{diagram:Simulation}
\end{figure}

This section explains our simulation scenario and illustrates some representative numerical results to compare the performance of different multiple-access techniques in terms of achievable spectral efficiency.
As shown in \figurename \ref{diagram:Simulation}, we consider the cell coverage consisting of a cell-center area and a cell-edge area. To facilitate the simulation setup, the cell-center area is a square with the side length of \SI{300}{\meter}. The cell-edge area starts from $(250\si{\meter},250\si{\meter})$ and terminates at $(500\si{\meter},500\si{\meter})$. The BS is located at the original point $(0,0)$ of the coordinate system, while a reflecting surface is installed at the center of the cell-edge area, with the coordinate $(375\si{\meter},375\si{\meter})$.  Half of the users distribute randomly over the cell-edge area while the other half of users distribute randomly over the cell-edge area. The break points of the three-slope model in (\ref{eqn:CostHataModel}) take values $x_0=10\mathrm{m}$ and $x_1=50\mathrm{m}$. The quantity $\mathcal{P}_0=140.72\mathrm{dB}$ with the carrier frequency of $1.9\mathrm{GHz}$, the BS antenna height of $15\mathrm{m}$, and the UE antenna height of $1.65\mathrm{m}$, while the standard derivation for shadowing is $\sigma_{sd}=8\mathrm{dB}$.
The maximum transmit power of BS is $P_d=20\mathrm{W}$ over a signal bandwidth of $B_w=20\mathrm{MHz}$, conforming with the practical 3GPP LTE specification. The variance of white noise is figured out by $\sigma_n^2=\kappa\cdot B_w\cdot T_0\cdot N_f$ with the Boltzmann constant $\kappa$, temperature $T_0=290 \mathrm{Kelvin}$, and the noise figure  $N_f=9\mathrm{dB}$.  Other simulation parameters are as follows: $N=200$, $N_b=16$, $L_0=\SI{-30}{\decibel}$, $\Gamma=5$, and $\alpha=2$.

\begin{figure*}[!t]
\centerline{
\subfloat[]{
\includegraphics[width=0.42\textwidth]{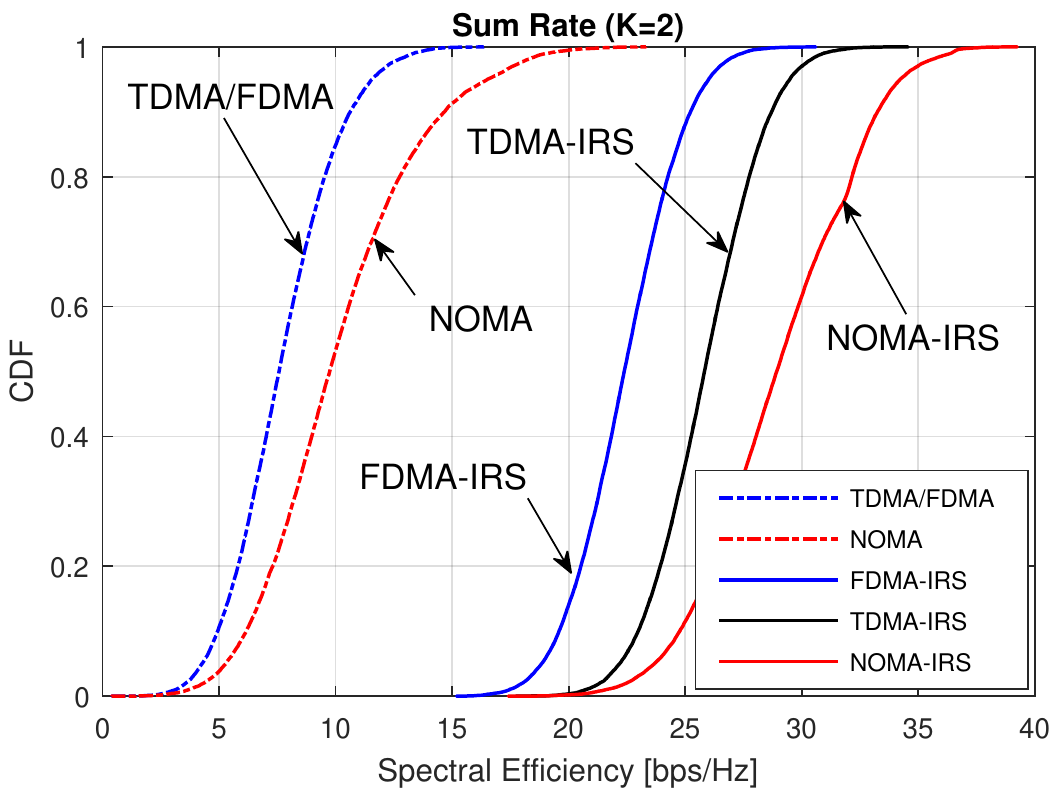}
\label{Fig_results1}
}
\hspace{10mm}
\subfloat[]{
\includegraphics[width=0.42\textwidth]{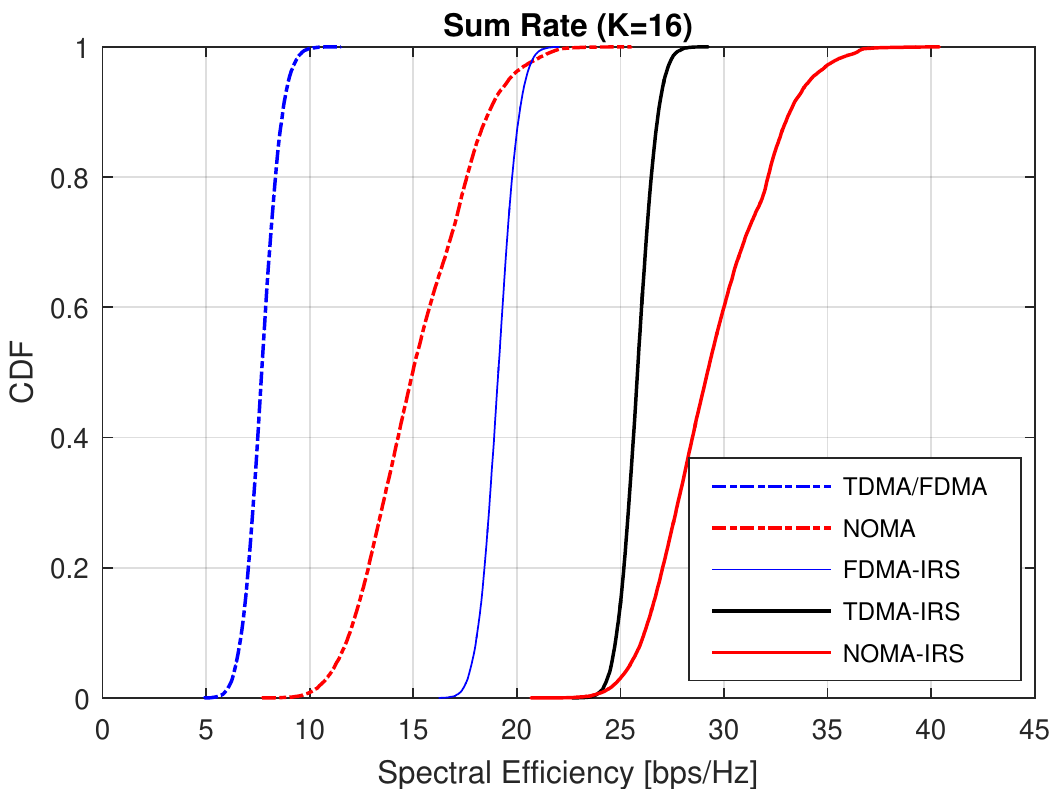}
\label{Fig_results2}
}}
\caption{Numerical results of different OMA and NOMA schemes in an IRS-aided multi-user MIMO system: (a) CDFs in terms of the sum spectral efficiency with two users, and (b) CDFs in terms of the sum spectral efficiency with sixteen users. }
\label{Fig_Result}
\end{figure*}

In \figurename \ref{Fig_results1}, we compare the cumulative distribution functions (CDFs) of the sum spectral efficiency achieved by the IRS-aided multi-user MIMO system with $K=2$ users, consisting of a cell-center user and a cell-edge user. Different schemes for comparison include: 1) TDMA without the aid of IRS, where the BS applies the MRT $\mathbf{w}_k^\star=\mathbf{f}_k^*/\|\mathbf{f}_k\|$ to achieve matched filtering in terms of the BS-UE direct link. Note that FDMA without the aid of IRS achieves the same performance, which is omitted in the figures for simplicity;  2) NOMA without the aid of IRS; 3) FDMA-IRS; 4) TDMA-IRS; and 5) NOMA-IRS. In our simulations, the number of iterations for alternating optimization is set to three, which is sufficient for the convergence of optimization.

The TDMA or FDMA scheme achieves the $95\%$-likely spectral efficiency, which is usually applied to measure the performance of cell-edge users, of \SI{4.26}{\bps\per\hertz^{}}, and the $50\%$-likely or median spectral efficiency of  \SI{7.57}{\bps\per\hertz^{}}. As we expected, NOMA is superior to OMA due to the full use of the time-frequency resource. To be specific, NOMA boosts the $95\%$-likely and $50\%$-likely rates to \SI{5.28}{\bps\per\hertz^{}} and \SI{9.74}{\bps\per\hertz^{}},  amounting to the rate growth of $24\%$ and $29\%$, respectively. It is observed that the application of IRS brings substantial performance improvement.
The $95\%$-likely and $50\%$-likely rates of FDMA-IRS grow to \SI{18.83}{\bps\per\hertz^{}} and \SI{22.38}{\bps\per\hertz^{}}, respectively, amounting to four-fold and three-fold spectral efficiencies compared to FDMA. Compared with TDMA-IRS, which achieves  $95\%$-likely spectral efficiency of \SI{22.18}{\bps\per\hertz^{}} and $50\%$-likely spectral efficiency of \SI{25.82}{\bps\per\hertz^{}}, there is a loss of approximately \SI{3.5}{\bps\per\hertz^{}}. That is because time-selective IRS elements can aid both users optimally in TDMA-IRS by dynamically changing the phase shifts in different slots, whereas the cell-center user in FDMA-IRS suffers from phase-unaligned reflected signals due to the lack of frequency-selective IRS reflection. As in the conventional systems, NOMA in an IRS-aided system still outperforms the two OMA schemes, achieving the best performance of \SI{23.67}{\bps\per\hertz^{}} and \SI{28.92}{\bps\per\hertz^{}}, respectively. In addition, we also illustrate the numerical results of these schemes in the case of $K=16$ users. As we can see in \figurename \ref{Fig_results2},  similar conclusions can be drawn from their performance comparison.

\section{Conclusions}
This paper developed and analyzed novel multiple-access schemes coined TDMA-IRS, FDMA-IRS, and NOMA-IRS for IRS-aided multi-user MIMO systems. Their design was based on an alternating method to jointly optimize active beamforming at the BS and passive reflection at the IRS. Theoretical analysis and numerical evaluation revealed that FDMA is inferior to TDMA in an IRS-aided system due to the lack of frequency-selective IRS elements. Meanwhile,  NOMA  still outperforms OMA as in a conventional system, with the price of complex signal processing for superposition coding and SIC. This paper will help to clarify which multiple-access technique is more favorable, and inspire the design of more efficient schemes for this emerging 6G paradigm.





%

\bibliographystyle{IEEEtran}
\bibliography{IEEEabrv,Ref_Globecom}

\end{document}